\documentclass{an}
\usepackage{graphicx}
\usepackage{times}
\usepackage{fancyhdr}
\begin{document}

\title{Autonomous software - myth or magic?}

\author{A.\ Allan\inst{1}, T.\ Naylor\inst{1}, E.S.\ Saunders\inst{1,2}}
\institute{School of Physics, University of Exeter, Stocker Road, Exeter, EX4 4QL, U.K.
\and
Las Cumbres Observatory, 6740 Cortona Dr. Suite 102, Santa Barbara, CA 93117, U.S.A.}

\date{Received $<$date$>$; accepted $<$date$>$; published online $<$date$>$}

\abstract{We discuss work by the eSTAR project which demonstrates a fully closed loop autonomous system for the follow up of possible micro-lensing anomalies. Not only are the initial micro-lensing detections followed up in real time, but ongoing events are prioritised and continually monitored, with the returned data being analysed automatically. If the ``smart software'' running the observing campaign detects a planet-like anomaly, further follow-up will be scheduled autonomously and other telescopes and telescope networks alerted to the possible planetary detection. We further discuss the implications of this, and how such projects can be used to build more general autonomous observing and control systems.
\keywords{telescopes, methods: observational, methods: miscellaneous, instrumentation: miscellaneous, astronomical databases: miscellaneous}}

\maketitle

\section{Introduction}

\begin{figure}
\resizebox{\hsize}{!}
{\includegraphics[]{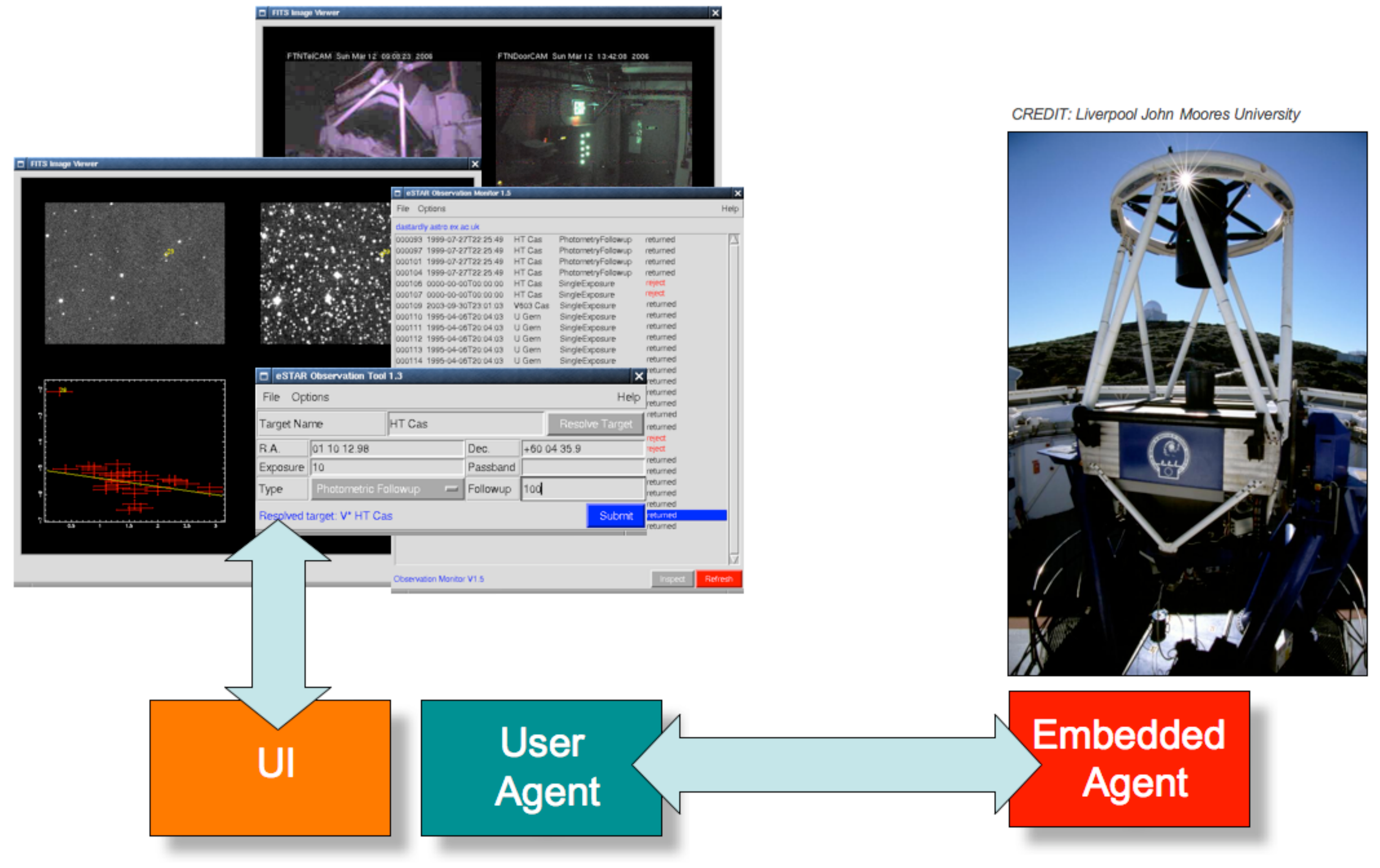}}
\caption{A simple block diagram showing how the user would, in the simplest case, make a direct observation request. Here an observation request is made by the user, and distributed by the user's agent to all agents embedded at telescopes on the network, who then score the request and return their evaluation of how well they could perform the observation to the user's agent. The user's agent then chooses the telescope best able to carry out the observation and places an observation request with that telescope. Once accepted, the observations are queued and carried out by the telescope with the data products being returned by the embedded agent to the user's agent. The agent may either then return these to the user directly, or possibly carry out autonomous follow-up depending on the results of the observation run.}
\label{fig:how_it_works}
\end{figure}

In the last few years the ubiquitous availability of high bandwidth networks has changed the way both robotic and non-robotic telescopes operate, with single isolated telescopes being integrated into expanding smart telescope networks that can span continents and respond to transient events in seconds. 

Two standards bodies, the IVOA\footnote{{\tt http://www.ivoa.net/}} and the HTN\footnote{{\tt http://www.telescope-networks.org/}}, have emerged in response to these changes, and from these two bodies two separate, but complimentary, standards are being developed in parallel.

The IVOA is working on the VOEvent standard which has been designed to transport timely information concerning transient events (White et~al. 2006). The eSTAR\footnote{{\tt http://www.estar.org.uk/}} project and the HTN have developed standards (e.g. Allan et~al. 2006a, Allan et~al. 2006b) which have been designed to allow the recipient of an event message to negotiate for, and obtain, follow-up observations to these reported events from a heterogeneous collection of networked telescopes.

\section{The eSTAR system}

The eSTAR Project (Allan et~al. 2004) was funded as part of the UK e-Science core programme to establish an intelligent robotic telescope network. It is a joint project between the Astrophysics Research Institute at Liverpool John Moores University and the Astrophysics Research Group of the School of Physics at the University of Exeter, in collaboration with the Joint Astronomy Centre (JAC) in Hawaii. 

Traditionally humans construct the complicated telescope schedules utilised by classical observing  using a multi-pass approach. The eSTAR project implements the collaborative agent paradigm (Wooldridge 2002), utilising the partial plan model, with a flat peer-to-peer network topology, which schedules a collection of distributed telescopes and attempts to imitate this approach.

\begin{figure}
\resizebox{\hsize}{!}
{\includegraphics[]{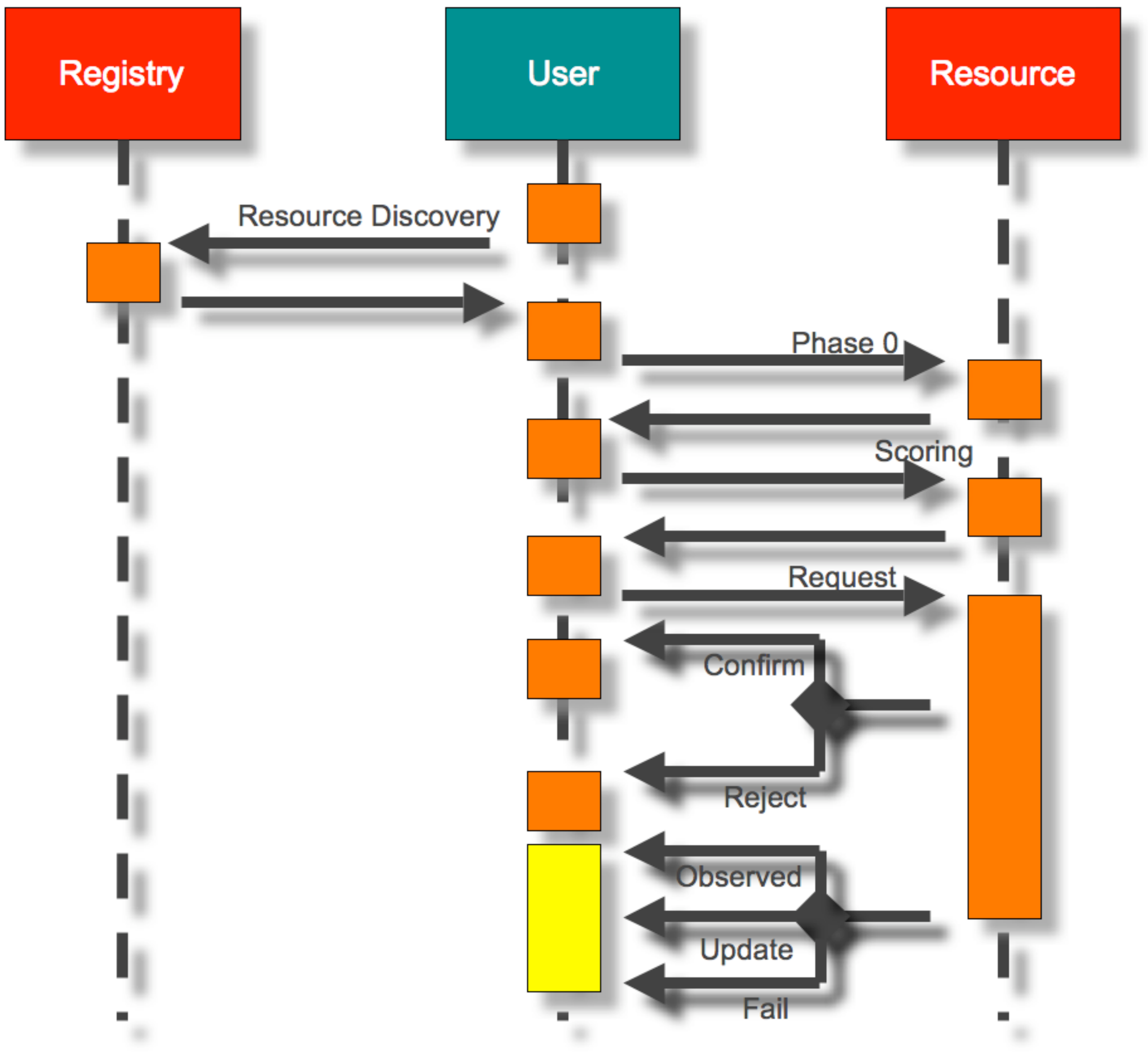}}
\caption{A diagram showing the message dialogue inside the eSTAR system during the course of an observation programmne. The message dialogue (Allan et al. 2006a) breaks down into five distinct phases or transactions. Firstly we carry out resource discovery to find out where the telescopes are, and what capabilities and services they provide. A preliminary exchange of requests, and corresponding scoring responses, then occurs to evaluate the ability of the telescope to perform the desired observations. An exchange of observation request and the corresponding confirmation or rejection of the observing request follows. The dialogue completes with the return of any data and final status of the observations to the user.}
\label{fig:htn_protocol}
\end{figure}

The project has developed the protocols, transport standards, and software (Allan et al. 2006b) to allow remote and geographically distributed telescopes to talk to one another, see Figure~\ref{fig:htn_protocol}. The project represents a ``turn-key'' system for autonomous observations of transient events and for long term monitoring campaigns which would otherwise be too onerous to handle manually (e.g. Saunders, Naylor, Allan 2006). 

In our architecture both the software controlling the science programme, and the software embedded at the telescope acting as a high-level interface to the native telescope control software, are thought of as agents. A negotiation takes place between these agents in which each of the telescopes bids to carry out the work, see Figure~\ref{fig:htn_protocol}, with the user's agent scheduling the work with the agent embedded at the telescope that promises to return the best result. 

This architectural distinction of viewing both sides of the negotiation as agents, and as equals, is crucial. Importantly this preserves the autonomy of individual telescope operators to implement scheduling of observations at their facility as they see fit, and offers adaptability in the face of asynchronously arriving data. For instance an agent working autonomously of the user can change, reschedule, or cancel queries, workflows or follow-up observations based on new information received.

However how well should one bit of software believe other bits of software, running at remote sites, when they tell it something is true? The eSTAR system fundamentally relies on the concept of scoring (Figures~\ref{fig:how_it_works} and~\ref{fig:htn_protocol}). If the agents embedded at the telescopes are returning unreliable scores, then the system is compromised. This scenario could occur for a number of reasons, e.g. overly-optimistic metrics at the telescope, unreliable hardware, or a period of changeable weather.

We are currently building software to keep track of scores returned from telescopes, correlating these with the data actually returned as a result of the observation, i.e. we are trying to evaluate in real time the accuracy of the telescope's scoring prediction. The agent running our science programme will use this to modify our decision making process in the future, hopefully weighting things towards the telescopes that can more accurately predict their chance of performing the requested observations. With a feedback mechanism to the embedded agents in place, we hope to show some sort of emerging social conventions (Shoham \& Tennenholtz 1997) between the negotiating pieces of software, where ``good'' behaviour is rewarded and ``bad'' behaviour is punished, will eventually emerge. As a result, fewer observing requests will be made to poorly performing telescopes.

\section{Hunting for exo-planets}

Microlensing is currently the fastest and cheapest way to search for cool planets. It is this technique (Horne 2008) that is being utilised by eSTAR and RoboNet-1.0\footnote{{\tt http://www.astro.livjm.ac.uk/RoboNet/}} to intensively monitor large numbers of Galactic Bulge microlensing events. The method is most sensitive to cool planets, 1--5 AU from the lens stars and is the only ground-based technique that is currently capable of discovering Earth-mass planets.

The project is attempting to maximise the planet discovery rate by optimally observing the known ongoing microlensing events. The frequency and length of the observations of each lensing event is determined by a number of factors, including he locations, availability and current conditions at the telescopes and the lensing strength of the event itself.

At the start of the 2007 bulge season the project began full autonomous follow-up of the micro-lensing events, with real time data reduction and analysis, and full reactive scheduling. Our system should detect a possible planetary anomaly in a microlensing light curve, and autonomously follow up and confirm this anomoly, while simultaneously notifying the community of a exo-planet candidate using the VOEvent network.

\subsection{First-look observations}

\begin{figure}
\resizebox{\hsize}{!}
{\includegraphics[]{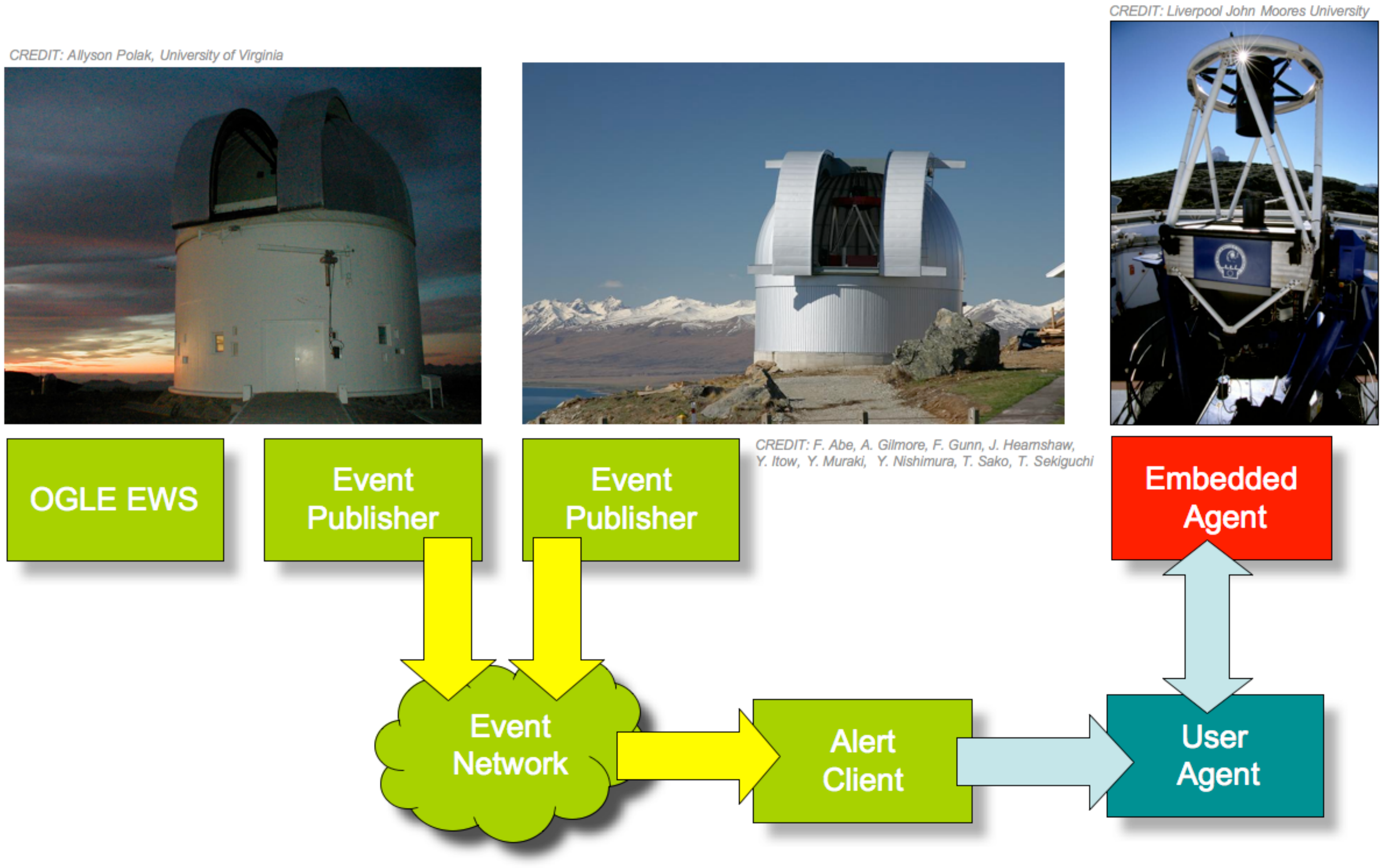}}
\caption{A block diagram showing how first look observations are made. A VOEvent message providing notification of a newly discovered microlensing event is generated, and received in Exeter by the event broker. The message is passed by the broker to subscribing clients (see Allan et al. 2008 for details). The alert client will compare the message to a pre-definied set of criteria, e.g. brightness, to see if if should be followed up. If the event is recommended for follow-up a request is placed with the user agent to obtain a short time series. In this scenario the user and graphical interface shown in Figure~\ref{fig:how_it_works} is replaced by the autonomous request made by the alert client, however in all other respects the process is the same as if the user placed the request directly with the agent.}
\label{fig:first_look}
\end{figure}

During the 2007 season first look observations, see Figure~\ref{fig:first_look}, were taken of all OGLE-EWS\footnote{{\tt http://www.astrouw.edu.pl/$\sim$ogle/ogle3/}} events. These were short time series taken in response to the VOEvent messages notifying the eSTAR system of the presence of a new microlensing event and served as a calibration for ongoing monitoring. During the upcoming 2008 season first look observation will also be taken for MOA\footnote{{\tt http://www.phys.canterbury.ac.nz/moa/}} events.

\subsection{Ongoing monitoring}
  
\begin{figure}
\resizebox{\hsize}{!}
{\includegraphics[]{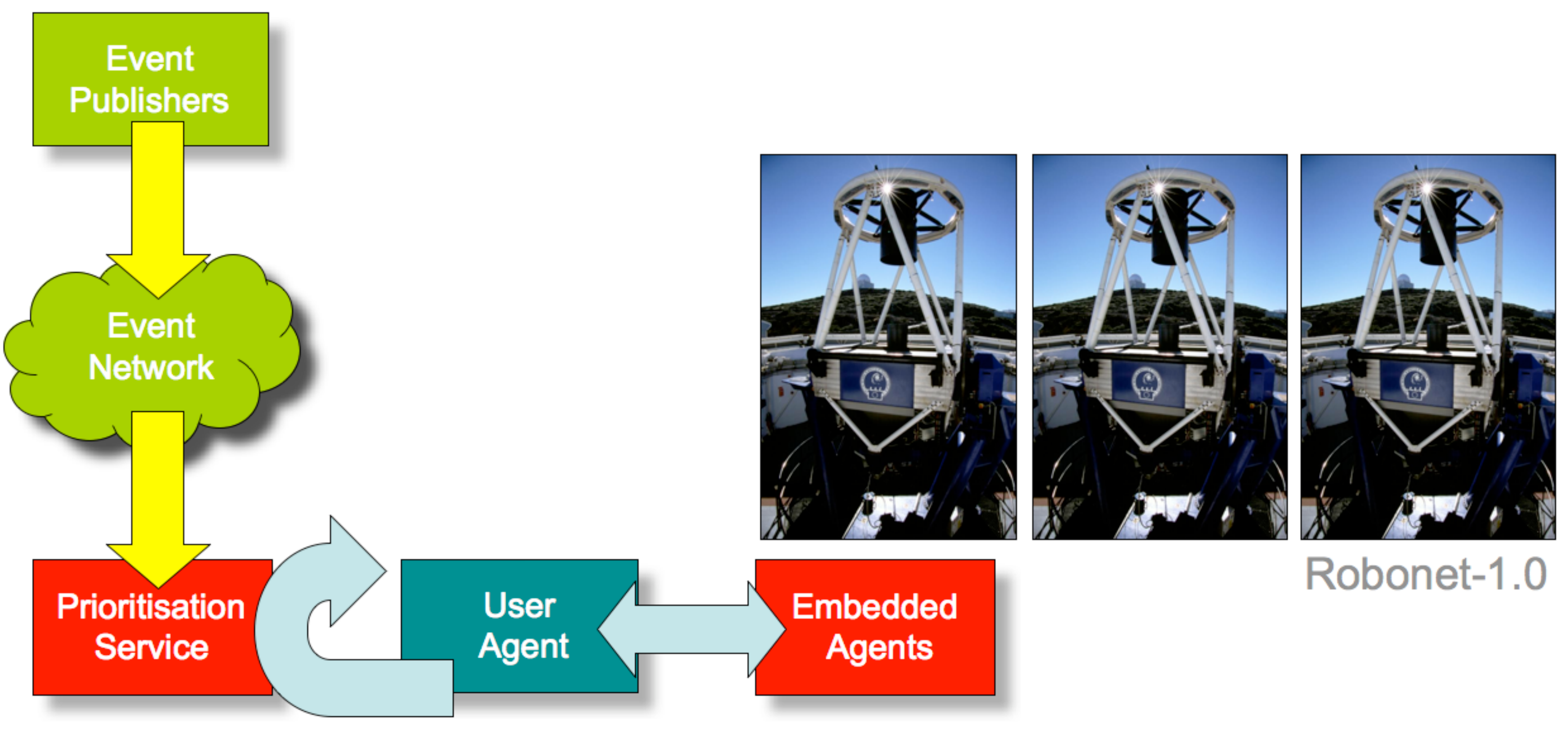}}
\caption{A block diagram showing how the monitoring of ongoing events is carried out. Here a prioritisation service keep track of all the active microlening events. This service is then polled periodically by the agent in charge of the science programme, and a list of observing requests generated. Things then progress as for Figure~\ref{fig:how_it_works} and a manually submitted request.}
\label{fig:ongoing_monitoring}
\end{figure}

Microlensing follow-up is a complex problem, requiring two stages of alert. First one must detect that a star is undergoing a microlensing event, and then determine that it is ``anomalous'', suggesting the presence of a planet. Then very frequent observations are required over periods of hours to days, see Figure~\ref{fig:ongoing_monitoring}. Speed of response is less important, but often several events are occurring at once, presenting a complex scheduling problem (Horne 2008) of maximising the chances of detecting a planet by applying the maximum coverage to the events currently in progress which are most likely to yield planets.

\subsection{Closing the loop}

The feedback capabilities enabled by the eSTAR network, which allows autonomous real time evaluation of data products received by the agent running the user's science products, either via the event network or from the agent's own observing requests, allows  ``on the fly'' decision making as to the nature of the follow-up. This in-built flexibility can prove to be a crucial resource for rapid transient follow-up, and is currently an ability unique to the eSTAR network.

Running inside the feedback loop is an anomaly detector (see Dominik et al. 2007) which provides  autonomous decision making capability, this allows us to build systems which will learn and adapt. With the eSTAR system ``feeding'' the anomaly detector with data in real time the software is used to provide an expert opinion, allowing the system itself to solve the distributed scheduling problem of where best to follow-up a potential exo-planetary detection entirely autonomously from the user.

\section{Conclusions}

We have proved that the eSTAR architectural approach works well for a diverse collection of telescopes. The potential to organically grow the eSTAR network to include new telescopes and VO enabled-databases is obvious, and in doing so we potentially open up new parameter spaces. The system is fully autonomous and is based on open standards, it represents a Òturn-keyÓ system operating in three modes; event driven, long term monitoring and reactive follow-up. It is a fully closed loop system and can operate without human intervention.

\acknowledgements  The authors would like to acknowledge the support of the DTI, EPSRC, PPARC and STFC which jointly fund the eSTAR project.

\end{document}